\documentclass[aps,prb,floatfix,twocolumn]{revtex4}
\usepackage[pdftex]{graphicx}
\usepackage{amsmath,amssymb,color,verbatim}
\usepackage{mdframed}
\usepackage{setspace}
\usepackage{yllmath}

\providecommand{\myheading}[1]{ \textbf{#1}: }

\begin{document}
\title{A general method for calculating lattice Green functions on the branch cut}
\author{Yen Lee Loh}
\affiliation{Department of Physics and Astrophysics, University of North Dakota, Grand Forks, ND  58202}
\date{2017-5-21} 
\begin{abstract}
We present a method for calculating the complex Green function $G_{ij} (\omega)$ at any real frequency $\omega$ between any two sites $i$ and $j$ on a lattice. 
Starting from numbers of walks on square, cubic, honeycomb, triangular, bcc, fcc, and diamond lattices,
we derive Chebyshev expansion coefficients for $G_{ij} (\omega)$.
The convergence of the Chebyshev series can be accelerated by constructing functions $f(\omega)$ that mimic the van Hove singularities in $G_{ij} (\omega)$ and subtracting their Chebyshev coefficients from the original coefficients.
We demonstrate this explicitly for the square lattice and bcc lattice.
Our algorithm achieves typical accuracies of 6--9 significant figures using 1000 series terms.
\end{abstract}
\maketitle

Consider the quantum mechanical tight-binding Hamiltonian
	\begin{align}
	\hat{H}
	&= t \sum_{\mean{ij}} \big(   \ket{i} \bra{j} + \ket{j} \bra{i} \big)
	\label{Hamiltonian}
	\end{align}
where $t$ is the hopping amplitude between nearest-neighbor sites $i$ and $j$ on a lattice.  Define the ``Greenian'' operator
	$
	\hat{G} (\omega)
	=  ( \omega -  \hat{H}  )^{-1}
	$
where $\omega  -  \hat{H} \equiv \omega \hat{1} - \hat{H}$ and $\hat{1}$ is the identity operator.
The matrix elements of the Greenian
	\begin{align}
	G_{ij} (\omega) 
	&=  \bra{i}  (\omega  -  \hat{H})^{-1}   \ket{j}
	\label{Gdefinition}
	\end{align}
are called the lattice Green function (LGF).  In this paper we consider $i$ and $j$ to be fixed, and we will often omit these indices for brevity.

Lattice Green functions are not limited to quantum mechanics, but arise frequently in many other areas of physics.  The ability to compute LGFs can be useful, for example, for simulations of Hubbard models\cite{bloch2008review} and for non-perturbative renormalization group studies of scalar boson models.\cite{caillol2012,caillol2013}
For $i=j$, $G_{ij} (\omega)$ can be expressed as closed forms in terms of named special functions (mainly elliptic integrals or generalized hypergeometric functions) for square, bcc, honeycomb, diamond, cubic, hypercubic, triangular, and fcc lattices.\cite{guttmann2010,maradudin1960,morita1971,katsura1971,joyce1972,joyce1994,joyce2002,joyce2001,joyce2003,delves2001,ray2014arxiv,schwalm1988,schwalm1992}  For $i \neq j$, $G_{ij} (\omega)$ can be expressed as closed forms for square, bcc,\cite{ray2014arxiv} honeycomb,\cite{ray2014arxiv} triangular,\cite{ray2014arxiv} kagome,\cite{ray2014arxiv} diced,\cite{ray2014arxiv} and cubic\cite{joyce2002} lattices; spatial recurrence relations exist but are often numerically unstable.\cite{morita1971recurrence,morita1975,berciu2009}
In this paper we develop a general numerical method applicable to LGFs for which no closed form is known.

\begin{figure}[!htb]
		\includegraphics[width=0.6\columnwidth]{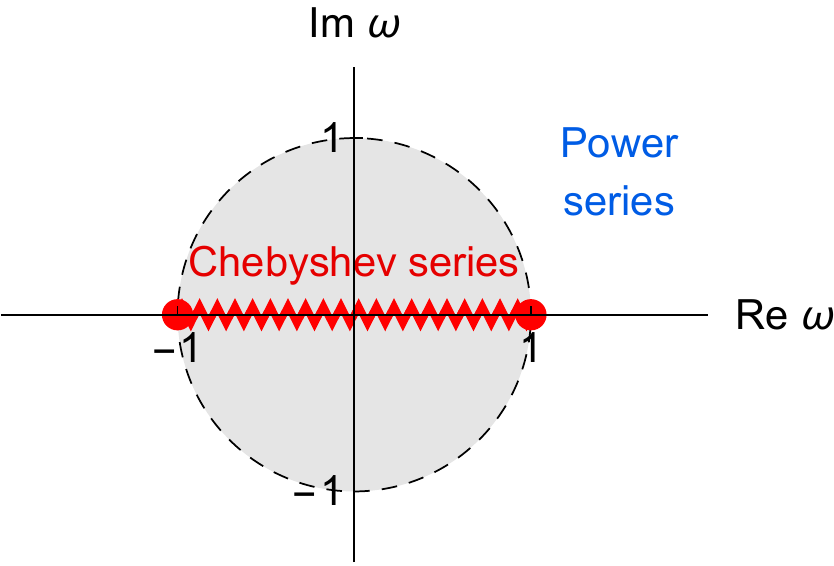}	
	\caption{
	\label{Argand}
		Structure of a lattice Green function $G_{ij} (\omega)$ in the complex plane.
		The power series converges outside the unit disk,
		whereas the Chebyshev series is applicable just above the branch cut.
	}
	\end{figure}

In the complex $\omega$ plane, the Green function has a branch cut running from $\omega_\text{min}$ to $\omega_\text{max}$ along the real axis, where $\omega_\text{min}$ and $\omega_\text{max}$ are the lowest and highest eigenvalues of $\hat{H}$.
Wherever $\omega$ occurs as an argument of a Green function, it is to be interpreted as including an infinitesimal imaginary shift, i.e., as $\omega + i 0^+$.
We will always scale $\hat{H}$
such that all its eigenvalues lie within the interval $[-1,1]$ (see Fig.~\ref{Argand}).  For the lattices treated in this paper, we ensure this by choosing $t=1/z$ where $z$ is the coordination number (the number of neighbors of each site).
\footnote{For a physical tight-binding model, the hopping amplitudes are negative, so it would be more appropriate to choose $t=-1/z$.}

It is well known that the lattice Green function $G_{ij} (\omega)$ can be written as an inverse power series about $\omega=\infty$, where the coefficient of the $\omega^{-n-1}$ term is related to the number of paths of length $n$ from site $i$ to site $j$.
\cite{maassarani2000,mamedov2008}
This series is useful for numerically evaluating the Green function outside the unit disk ($\abs{\omega} > 1$), and sometimes to evaluate the quantities $G(\pm 1)$, which are known as Watson integrals.  (See Ref.~\onlinecite{guttmann2010} for a review.)
However, for certain applications, one requires values of the Green function ``on the cut'' (i.e., on either side of the branch cut), where the power series diverges.  This is a more difficult problem.  

The most direct approach is to write the Green function as a $d$-dimensional integral over the Brillouin zone, typically of the form
	\begin{align}
	G (\omega, \rrr) 
	&=\int_\text{BZ} \frac{d^d q}{(2\pi)^d}~
		\frac{\exp i\qqq\cdot\rrr}{\omega - \displaystyle\sum_{\pmb\delta} e^{i\qqq\cdot\pmb\delta} }
		.
		\label{BZIntegral}
	\end{align}
This expression does not lend itself well to numerical implementation because it involves high-dimensional integration of singular integrands.\cite{berciu2010} 
In previous papers \cite{loh2011hlgf,loh2013hlgf} we presented an approach for calculating Green functions accurately ($>12$ s.f.) based on time-frequency Fourier transformation and contour integration techniques.  Unfortunately, that approach relies on a factorization property of hypercubic lattice Green functions in the time domain, and we have not been able to generalize it to non-hypercubic lattices.  

Some of the literature focuses on finding ordinary differential equations that are satisfied by each lattice Green function.\cite{guttmann2010}  This does not appear to help directly with evaluating the LGFs on the cut.  

The recursion method and the continued-fraction method\cite{berciu2010,moller2012} can be used to evaluate LGFs on the cut, although accuracy appears to be limited to 2--6 decimal places.

In this paper we evaluate LGFs by converting their power series into Chebyshev series.  This amounts to analytic continuation from the region $\abs{\omega} > 1$ to the region $\abs{\omega} \leq 1$, as depicted in Fig.~\ref{Argand}.

\section{Counting walks on lattices\label{secCombinatorics}}
Tables~\ref{Wformulas} and \ref{Wvalues} show combinatorial formulas and explicit values for the numbers of polygons (closed walks) on various lattices.  Most of these results are well known in the literature.\cite{guttmann2010,domb1960,joyce1972,joyce1994,joyce2001,joyce2002,joyce2003,bailey2008} 
We reproduce them here for the reader's convenience, organized to reveal that the formulas fall into various families: the bcc family (1D chain, 2D square, 3D body-centered cubic); honeycomb family (2D honeycomb, 3D diamond); cubic family (1D chain, 2D square, 3D cubic, 4D hypercubic); and triangular family (2D triangular, 3D face-centered cubic).

Table~\ref{Wformulas} also gives formulas for the numbers of open walks beginning at the origin and ending at position $\rrr=(x,y,z)$ on the lattice.  We outline the derivations below.

\myheading{BCC family} 
Consider a walk on a bcc lattice.  
Let $n_x$, $m_x$, $n_y$, $m_y$, $n_z$, and $m_z$ be the numbers of steps in the $\pm x$, $\pm y,$ and $\pm z$ directions.  
Let $n$ be the total number of steps.
During each step, the walker moves \emph{simultaneously} in the $\pm x$, $\pm y,$ and $\pm z$ directions by either $+1$ or $-1$ units.
Thus $n = n_x + m_x = n_y + m_y = n_z + m_z$.
Let the net displacements in each direction be $x = n_x - m_x$, $y = n_y - m_y$, and $z = n_z - m_z$.
Then $n_x = \frac{n+x}{2}$, $n_y = \frac{n+y}{2}$, and $n_z = \frac{n+z}{2}$.  Thus the number of walks of length $n$ with total displacement $(x,y,z)$ is
	\begin{align}
	W_{xyzn}^\text{bcc}
	&=
		\binom{n}{\frac{n+x}{2}}
		\binom{n}{\frac{n+y}{2}}
		\binom{n}{\frac{n+z}{2}}
	.
	\end{align}
This derivation can easily be generalized to a $d$-dimensional bcc lattice.

\myheading{Cubic family} 
Consider a walk on a cubic lattice.  Suppose the numbers of steps in the $\pm x$, $\pm y,$ and $\pm z$ directions are given by the six integers $n_x$, $m_x$, $n_y$, $m_y$, $n_z$, and $m_z$.  The number of such walks is given by the multinomial coefficient
	\begin{align}
	\frac{n!}{ n_x!~ m_x!~ n_y!~ m_y!~ n_z!~ m_z!}
	.
	\end{align}
The total number of steps is $n = n_x + m_x + n_y + m_y + n_z + m_z$ and the net displacements in each direction are $x = n_x - m_x$, $y = n_y - m_y$, and $z = n_z - m_z$.  Let $j=m_x$, $k=m_y$, and $l=m_z$.  Then $n_x = j+x$, $n_y = k+y$, and $n_z = l+z$.  Furthermore, $n=2(j+k+l)+x+y+z$.  Thus the total number of walks of length $n$ with total displacement $(x,y,z)$ is
	\begin{align}
	W_{xyzn}^\text{cubic}
	&=
		\sum_{j=0}^{s}
		\sum_{k=0}^{s-j}
		\frac{n!} { j!~ (j+x)!~  k!~ (k+y)!~  l!~ (l+z)! }
	\end{align}
where $s \equiv \frac{n-x-y-z}{2}$ and $l \equiv s - j - k$.  
This derivation is easily generalized to a $d$-dimensional hypercubic lattice.
	
\begin{figure}[!bth]
		\includegraphics[width=0.9\columnwidth]{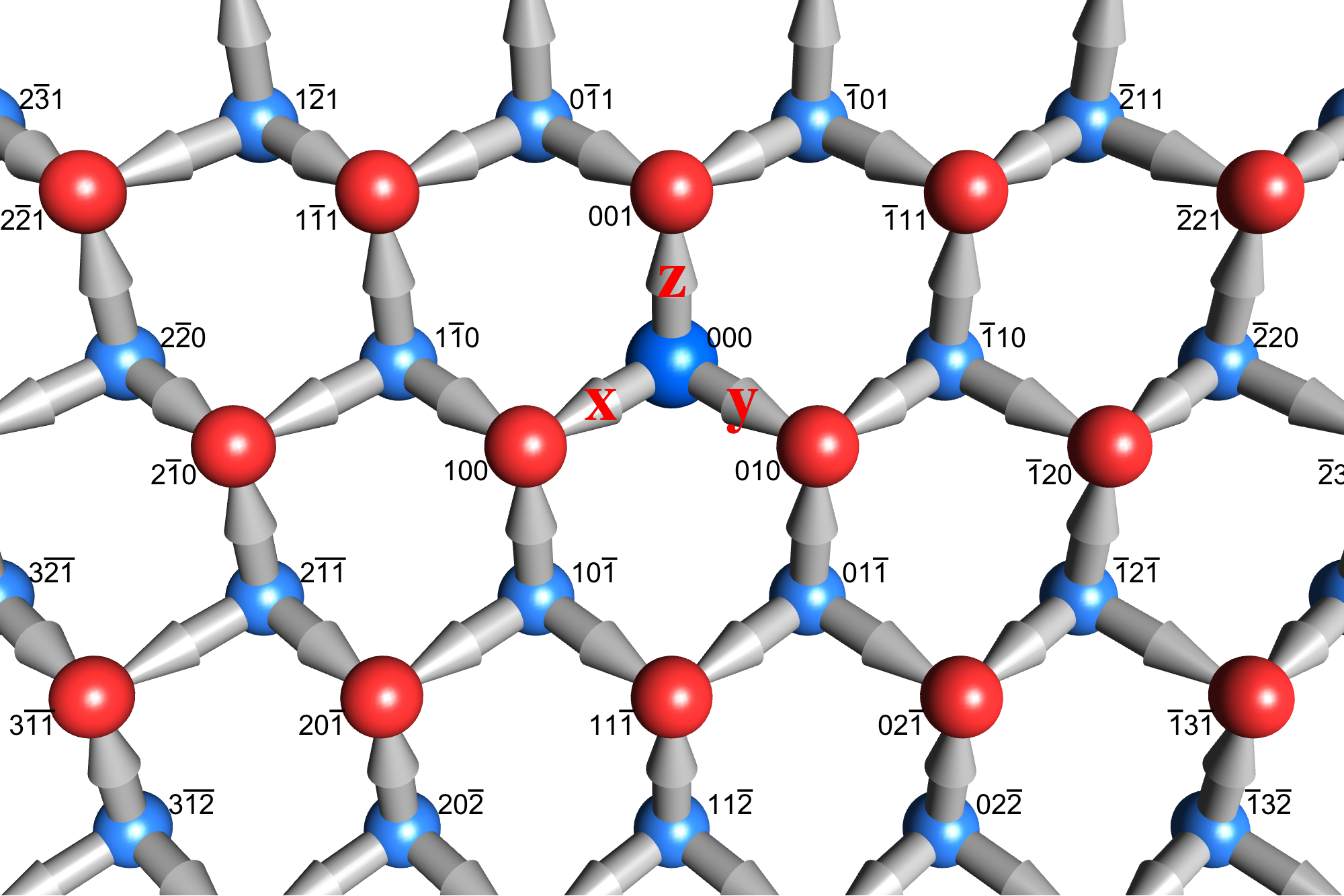}	
	\caption{
		\label{honeycomb}
		(Color online)
		Visualizing the honeycomb lattice as the projection of two (111) planes of a cubic lattice.
		The A sublattice (blue spheres) has coordinates $(x,y,z)$ satisfying $x+y+z=0$.
		The B sublattice (red spheres) has coordinates such that $x+y+z=1$.
	}
	\end{figure}

\myheading{Honeycomb family} 
The honeycomb lattice can be viewed as the projection of a puckered subset of a cubic lattice, as shown in Fig.~\ref{honeycomb}.  
Consider a walk starting at the origin.
Suppose the numbers of steps in the $\pm x$, $\pm y,$ and $\pm z$ directions are given by the six integers $n_x$, $m_x$, $n_y$, $m_y$, $n_z$, and $m_z$.
The total number of steps is $n = n_x + m_x + n_y + m_y + n_z + m_z$ and the net displacements in each direction are $x = n_x - m_x$, $y = n_y - m_y$, and $z = n_z - m_z$. 
Let $j=m_x$, $k=m_y$, and $l=m_z$.  Then $n_x = j+x$, $n_y = k+y$, and $n_z = l+z$.

On the odd-numbered steps of the walk, the walker can only travel in the $+x$, $+y$, or $+z$  directions.
Likewise, on even-numbered steps, the walker can only travel in the $-x$, $-y$, or $-z$  directions.

If $n$ is even, then there must be exactly $n/2$ steps along ``positive'' directions, and $n/2$ steps  along ``negative'' directions.  So $n/2 = n_x + n_y + n_z = m_x + m_y + m_z$.  Thus the number of walks of length $n$ is the number of permutations of step displacement vectors along odd-numbered steps, times the number of ways of permutations for even-numbered steps:
	\begin{align}
	W_{xyzn}^\text{hon}
	&=
		\sum_{j=0}^{n/2}
		\sum_{k=0}^{n/2-j}
		\frac{(n/2)!} { j!~ k!~ l!}
		\frac{(n/2)!} { (j+x)!~ (k+y)!~ (l+z)!}
	\end{align}
where $l \equiv n/2 - j - k$, and it is assumed that $x+y+z=0$.

If $n$ is odd, then there are $(n+1)/2$ positive steps and $(n-1)/2$ negative steps, so
	\begin{align}
	W_{xyzn}^\text{hon}
	&=
		\sum_{j=0}^{\frac{n-1}{2}}
		\sum_{k=0}^{\frac{n-1}{2}-j}
		\frac{(\frac{n+1}{2})!} { j!~ k!~ l!}
		\frac{(\frac{n-1}{2})!} { (j+x)!~ (k+y)!~ (l+z)!}
	\end{align}
where $l \equiv (n-1)/2 - j - k$.

Walks on a diamond lattice can be counted in a similar fashion.

\myheading{Triangular family} 
If one starts at the origin of the honeycomb lattice and performs two successive hops, there are three paths that return to the origin, and one path to each of the 6 A sites surrounding the origin.  Thus
$
	6\hat{H}^\text{tri}
	=
		(3\hat{H}^\text{hon})^2
	-	3(\hat{1})
$.
Therefore the number of paths on a triangular lattice of length $n$ with displacement $(x,y,z)$ (where $x+y+z=0$), using the same coordinate scheme, is
	\begin{align}
	W^\text{tri}_{xyzn} 
	&=\bra{\rrr} (6\hat{H}^\text{tri})^n \ket{\0}
		\nonumber\\
	&=\bra{\rrr} 
		\sum_{j=0}^n \binom{n}{j}  (3\hat{H}^\text{hon})^{2j} (-3)^{n-j} 
		\ket{\0}
		\nonumber\\
	&=
		\sum_{j=0}^n \binom{n}{j}  (-3)^{n-j}  W^\text{hon}_{x,y,z,2j} 
	.
	\end{align}
For closed walks $(x=y=z=0)$ this reduces to the formula for $W^\text{tri}_n$ shown in Table~\ref{Wformulas}, derived in Ref.~\onlinecite{guttmann2010} using a different approach.  

Similarly, using 
$
	12\hat{H}^\text{fcc}
	=
		(4\hat{H}^\text{diam})^2
	-	4(\hat{1})
$,
one obtains 
	\begin{align}
	W^\text{fcc}_{uvwsn} 
	&=
		\sum_{j=0}^n \binom{n}{j}  (-4)^{n-j}  W^\text{diam}_{u,v,w,s,2j} 
	\end{align}
in terms of the $(u,v,w,s)$ coordinates for diamond and fcc lattices embedded in a 4D grid.  The ``physical'' 3D Cartesian coordinates $(x,y,z)$ are given by the projection
	\begin{align}
	x &= u + v - w - s,	\\
	y &= u - v + w - s,	\\
	z &= u - v - w + s.	
	\end{align}

In the above discussion we have derived the $W_n$ using combinatorics.  There are other methods to obtain $W_n$, such as contour integration techniques.\cite{ray2014arxiv}

%
%

\section{Basic approach for calculating Green functions\label{secBasics}} 

\myheading{Power moments} 
Starting from the definition of $G(\omega)$, Eq.~\eqref{Gdefinition}, and expanding in powers of $\hat{H}$ shows that the Green function can be written in an inverse power series
	\begin{align}
	G(\omega) 
	= \sum_{n=0}^\infty  \frac{\bra{i}  \hat{H}^n  \ket{j}}{\omega^{n+1}} 
	.
	\label{GPowerSeries}
	\end{align}
From Eq.~\eqref{Hamiltonian} it is easy to see that $\bra{i}  \hat{H}^n  \ket{j} = t^n W_n = z^{-n} W_n$, 
where $W_n$ is the number of walks of length $n$ that begin at site $i$ and end at site $j$.  
As discussed earlier, Eq.~\eqref{GPowerSeries} converges only for $\abs{\omega} > 1$.

\myheading{Chebyshev polynomials} 
The Chebyshev polynomials of the first and second kinds are defined as
	\begin{align}
	T_n     (\cos \theta) &= \cos n\theta, \\
	U_{n-1} (\cos \theta) &= \sin (n\theta) / \sin\theta.
	\end{align}
They satisfy orthogonality and completeness relations
	\begin{align}
	\int_{-1}^1 dx~   \phi_n(x) \phi_{n'}(x) &= \delta_{nn'}
	\label{ChebyshevOrthogonality}
		, \\
	\sum_{n=0}^\infty \phi_n(x) \phi_n  (x') &= \delta(x-x'),
	\label{ChebyshevCompleteness}
	\end{align}
where $\phi_n(x) = \sqrt{ \frac{2-\delta_n}{\pi \sqrt{1-x^2}} }  T_n(x)$.  Here $\delta_{nn'}$ is the Kronecker delta function and $\delta_n \equiv \delta_{n,0}$.  The Chebyshev polynomials can be written in terms of monomials as
	\begin{align}
	T_n (x) &= \sum_{k=0}^n a_{nk} x^k	
	\label{ChebyshevsFromPowers}
		,\\
	a_{00} &= 1		\nonumber
		,\\
	a_{nk} &= (-1)^{(n-k)/2} n \frac{(n+k-2)!!}{k! (n-k)!!}
	\label{ankCoefficients}
	\end{align}
where $n=0,1,2,\dotsc,\infty$; $k=0,1,2,\dotsc,n$; and $n-k$ is even.  The first few coefficients $a_{nk}$ are shown in Table~\ref{ank}.  In \emph{Mathematica} we have found it fastest to to evaluate $a_{nk}$ using the recursion $a_{nk}=2a_{n-1,k-1}-a_{n-2,k}$ for $n\geq 2$.

The functions $T_n(x)/\sqrt{1-x^2}$ and $\pi U_{n-1}(x)$ are Hilbert transforms of each other; that is, they obey Kramers-Kronig relations:
	\begin{align}
	\int_{-1}^1 d\nu~ 
		\frac{T_n(\nu)}{(\nu-\omega) \sqrt{1-\nu^2} } 
		= \pi U_{n-1} (\omega)
		.
	\label{ChebyshevKK}
	\end{align}

\myheading{Chebyshev moments} 
Define the Chebyshev moment $g_n$ to be the matrix element of the Chebyshev polynomial of the Hamiltonian operator,
	\begin{align}
	g_n
	=\bra{i}  T_n(\hat{H})  \ket{j} 	.
	\end{align}
Using Eq.~\eqref{ChebyshevsFromPowers} we obtain
	\begin{align}
	g_n
	=\sum_{k=0}^n a_{nk} z^{-k} W_k
	\label{ChebyshevMomentsFromPowerMoments}
	\end{align}
where the coefficients $a_{nk}$ are given by Eq.~\eqref{ankCoefficients}.
Tables~\ref{ChebyshevMomentFormulas} and \ref{ChebyshevMomentValues} give formulas and values for Chebyshev moments on various lattices.

\myheading{Spectral function} 
Define the spectral function $g(\omega)$ as the matrix elements of a Dirac delta function,
	\begin{align}
	g(\omega) 
	=\bra{i}  \delta(\omega - \hat{H})  \ket{j}.
	\label{spectrum}
	\end{align}
This is a generalization of the local density of states.
Expanding the Dirac delta function using the Chebyshev polynomial completeness relation, Eq.~\eqref{ChebyshevCompleteness}, we see that the spectral function is the sum of an infinite series of  Chebyshev polynomials weighted by the Chebyshev moments,
	\begin{align}
	g(\omega) 
	=\frac{1}{\pi\sqrt{1-\omega^2}} \sum_{n=0}^\infty  (2-\delta_n) T_n(\omega)  g_n
	.
	\label{gSeries}
	\end{align}

\myheading{Green function} 
The spectral function and the Green function are related by $g(\omega)=-\frac{1}{\pi} \Im G(\omega+i 0^+)$ and $G(\omega) = \int_{-1}^1 d\nu~ \tfrac{1}{\omega - \nu + i 0^+} g(\nu)$.  Thus
	\begin{align}
	G(\omega)
	&= \int_{-1}^1 d\nu~ \left[ \tfrac{1}{\omega - \nu} - i \pi \delta(\omega - \nu) \right] g(\nu)
	.
	\end{align}
From Eq.~\eqref{gSeries} and Eq.~\eqref{ChebyshevKK} we then obtain
	\begin{align}
	G(\omega) 
	&= -\sum_{n=0}^\infty  (2-\delta_n) g_n
		\left[ U_{n-1} (\omega) + \frac{i  T_n (\omega) }{\sqrt{1-\omega^2}}   \right]
	.
	\label{GSeries}
	\end{align}

\myheading{Basic algorithm} 
Our approach for numerical computation of Green function may now be outlined as follows:
\begin{enumerate}
\item Calculate $\{W_n\}$ using combinatorial formulas.
\item Calculate $\{g_n\}$ using Eq.~\eqref{ChebyshevMomentsFromPowerMoments}.
\item Calculate $G(\omega)$ using Eq.~\eqref{GSeries}.
\end{enumerate}

\begin{figure}[!bth]
		\includegraphics[width=0.9\columnwidth]{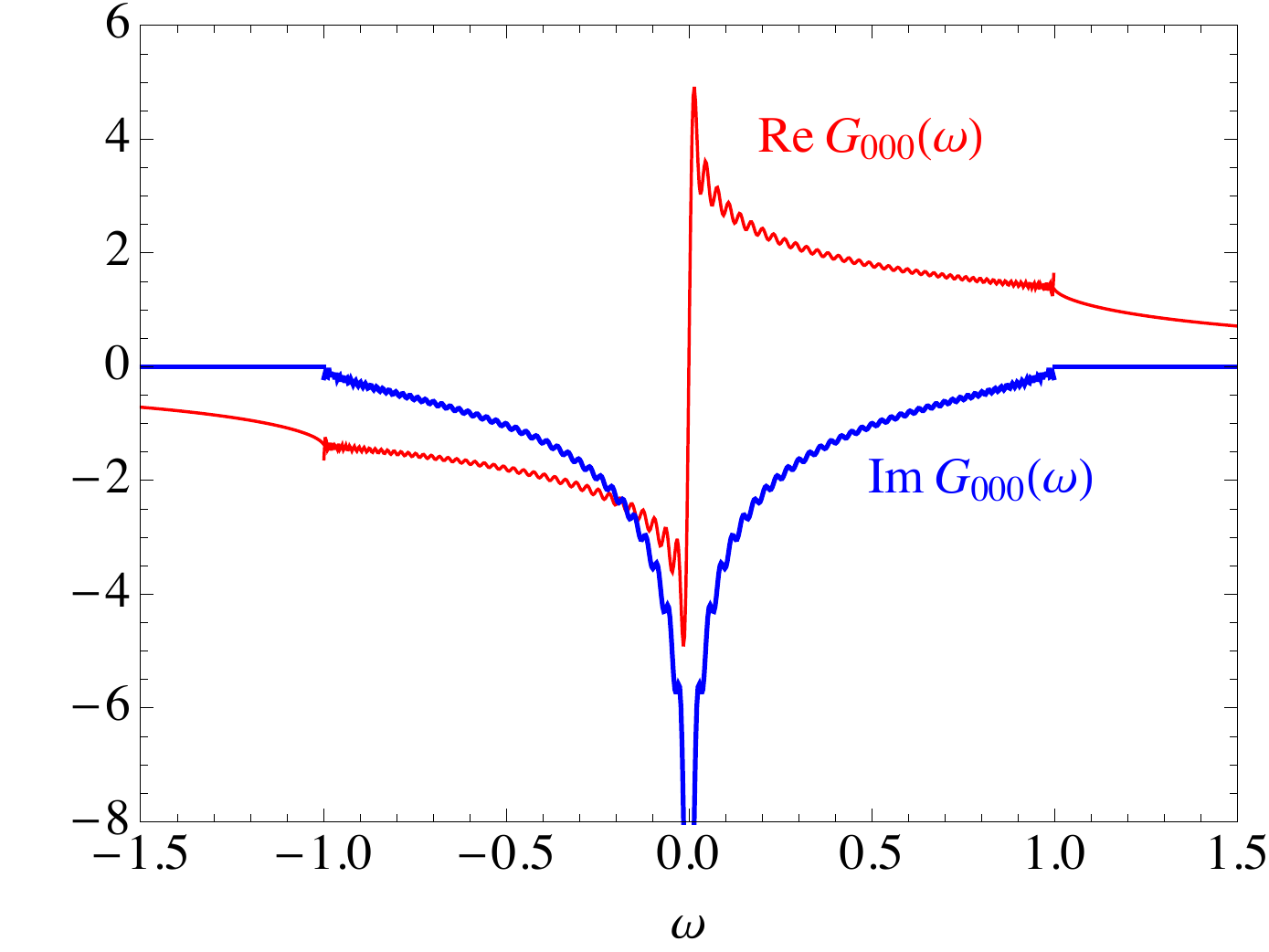}	
		\includegraphics[width=0.9\columnwidth]{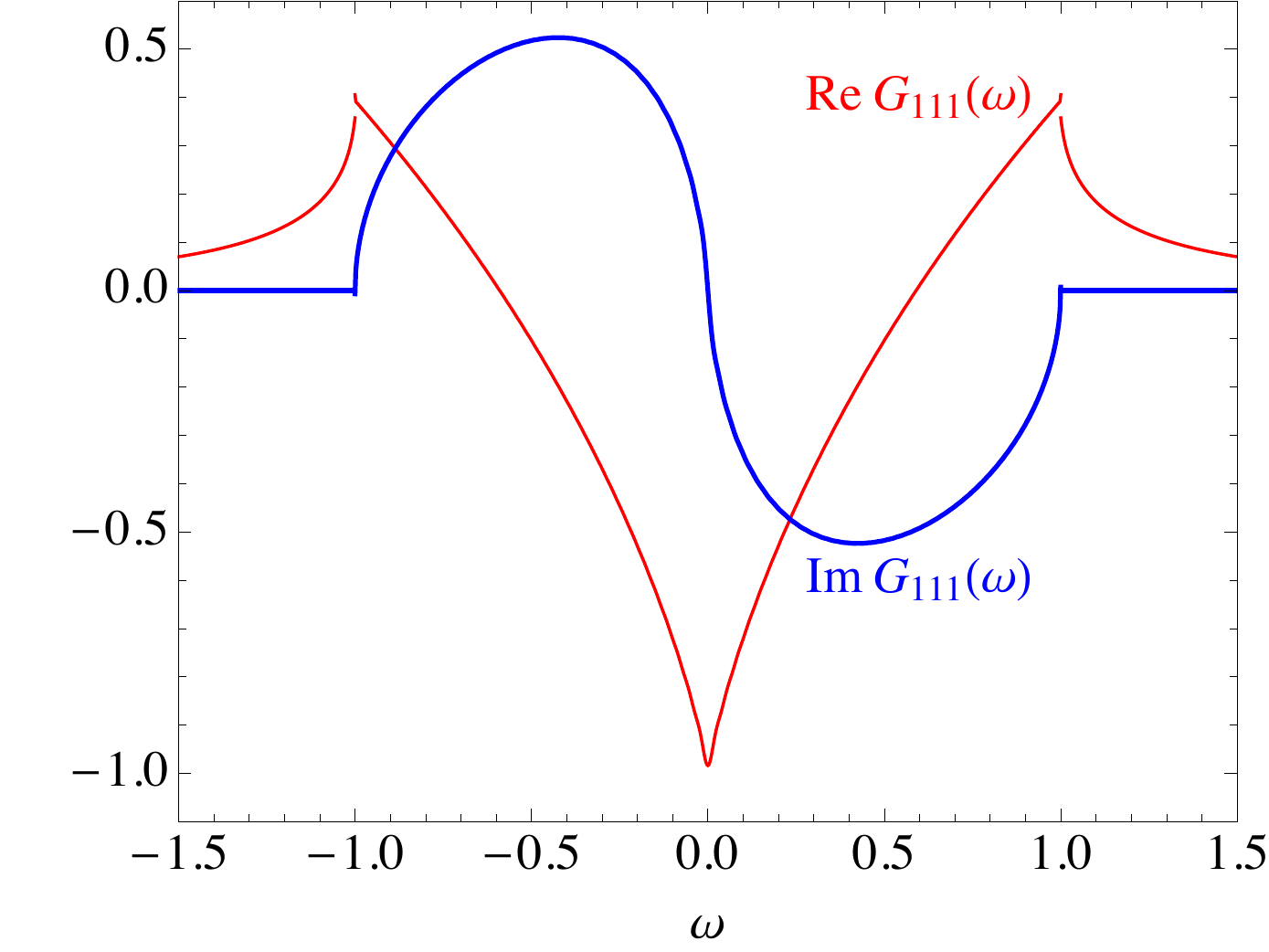}	
		\includegraphics[width=0.9\columnwidth]{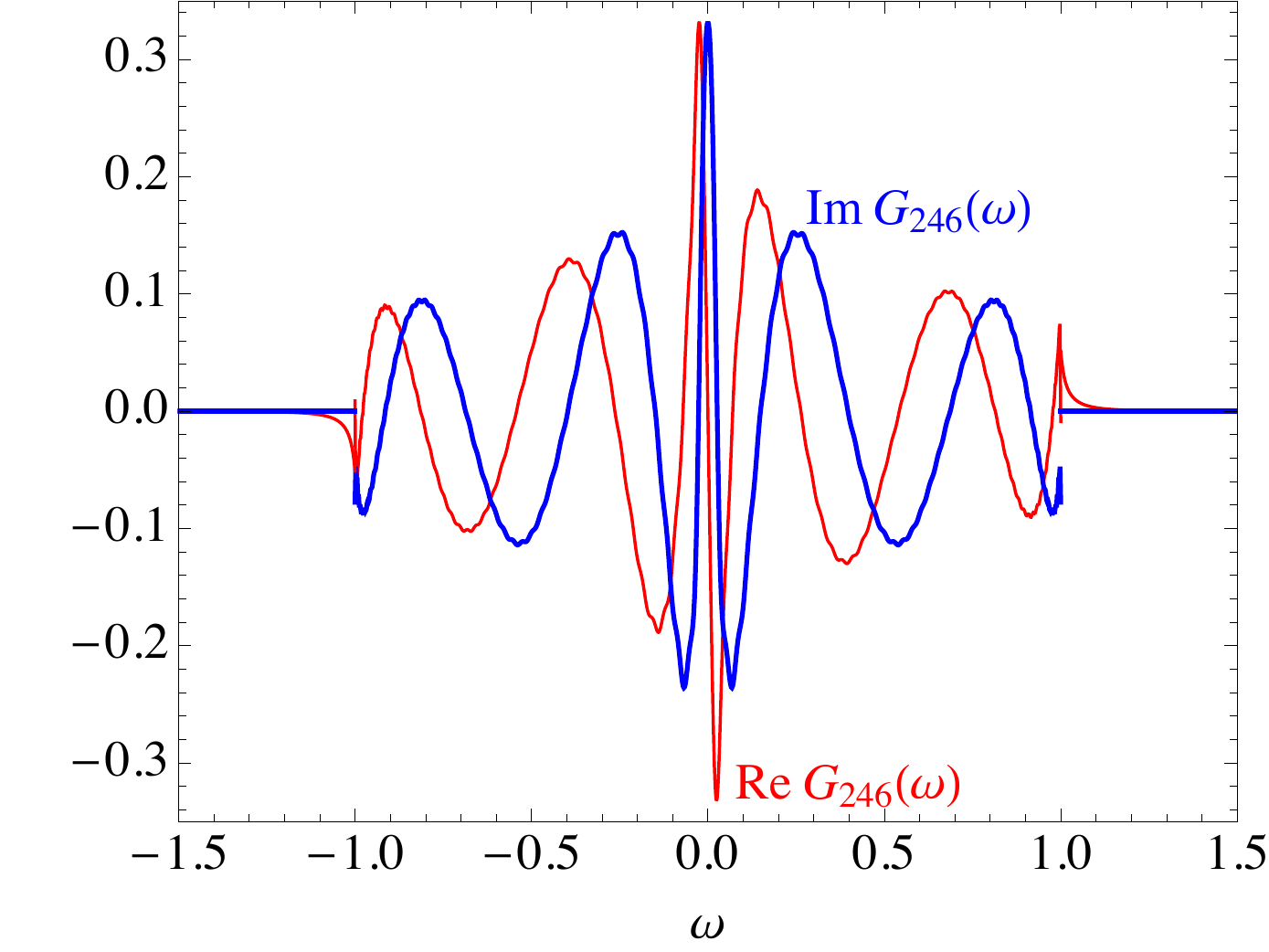}		
	\caption{
	\label{BCC-Gw}
		Real and imaginary parts of the bcc lattice Green function $G_{xyz}(\omega)$
			for various displacements $(x,y,z)$,
			computed using the methods of Sec.~\ref{secBasics}.
		For $\abs{\omega} > 1$ the Green function is computed using
			the inverse power series, Eq.~\eqref{GPowerSeries}, truncated at $n=200$.
		For $\abs{\omega} < 1$
			we use the Chebyshev series truncated at $n=200$,
			which is equivalent to a Fourier series using a rectangular window function.
			Gibbs oscillations are visible near $\omega=0$ and $\omega=\pm 1$.
		Truncation error produces visible discrepancies between 
			the power series and Chebyshev approximations at $\omega=\pm 1$.
	}
	\end{figure}

\myheading{Illustration}
Figure~\ref{BCC-Gw} shows the bcc lattice Green function $G_{\mathbf{i}\mathbf{j}} (\omega) \equiv G_{xyz} (\omega)$ for various displacements ${\mathbf{j}-\mathbf{i}} = (x,y,z)$, computed using this approach.

\myheading{Practical considerations} 
The sum in Eq.~\eqref{ChebyshevMomentsFromPowerMoments} involves large cancellations between terms, so it must be calculated using exact arithmetic.  Rewriting the sum as
	\begin{align}
	z^n g_n
	=\sum_{k=0}^n a_{nk} z^{n-k} W_k
	\end{align}
allows us to perform the computation exactly using integer arithmetic.  

Computing $g_n$ for $n=0,1,2,3,\dotsc,N$ directly from values of $W_k$ would require at least $O(N^2)$ time.  
By using expressions for $W_k$ we can derive expressions for $g_n$ that may be quicker to evaluate.  
See Table~\ref{ChebyshevMomentFormulas}.

The sum has an infinite number of terms.  If we truncate the series after $N$ terms, we necessarily introduce some error.  The Chebyshev series on $\omega\in [-1,1]$ is equivalent to a Fourier series on $\cos\omega\in [0,\pi]$, and so truncation error takes the form of Gibbs oscillations.  These can be mitigated by multiplying $g_n$ by an appropriate window function such as a Kaiser-Bessel window\cite{loh2014,karki2016}
before inserting it into Eq.~\eqref{GSeries}:
	\begin{align}
	G(\omega) 
	&= -\sum_{n=0}^L  (2-\delta_n) f^\text{win}_n g_n
		\left[ U_{n-1} (\omega) + \frac{i T_n (\omega)}{\sqrt{1-\omega^2}}   \right]
	,
		\nonumber\\
	f^\text{win}_n 
	&= \frac{I_0 \left( \beta\sqrt{1 - (\frac{n}{L-1})^2} \right)   }{ I_0(\beta)  }
	\label{GWindowed}
	\end{align}
where $\beta$ is a tuning parameter.
Nevertheless, accuracy is still limited.

%
%

\section{Van Hove singularities \label{secSingularities}} 
As illustrated in Fig.~\ref{BCC-Gw}, lattice Green functions contain van Hove singularities, which are difficult to expand in terms of Chebyshev polynomials.  They cause the Chebyshev coefficients $g_n$ to decay slowly as some inverse power of $n$.  We can accelerate the convergence of the series by identifying the functional forms of the singularities and subtracting the corresponding tails from the sequence of Chebyshev coefficients.

First let us review the theory of van Hove singularities.  The density of states on a lattice is 
	\begin{align}
	g(\omega)
	=\int_\text{BZ} \frac{d^dk~}{(2\pi)^d} \delta \big(\omega - \vare(\kkk)\big)
	\end{align}
where $\vare(\kkk)$ is the dispersion relation and $\delta$ is the $d$-dimensional Dirac delta function.  Roughly speaking, this may be written as
	\begin{align}
	g(\omega)
	=\int \frac{d^{d-1}k_\perp}{(2\pi)^d}~  \frac{1}{\abs{\nabla\vare(\kkk)} }
	.
	\end{align}
When the wavevector $\kkk$ approaches a critical point $\kkk_0$ where $\nabla\vare=\0$, the integrand diverges, producing a van Hove singularity in $g(\omega)$ at $\omega=\vare_0=\vare(\kkk_0)$.  The dispersion relation may be expanded as a power series about the critical point.  Typically
	\begin{align}
	g(\omega)
	=\int \frac{d^dq}{(2\pi)^d}~ 
		\delta \big(\omega - \vare_0 - \frac{\lambda_1 {q_1}^2 + \dotsc + \lambda_d {q_d}^2}{2} \big)
	\label{DispersionHessianExpansion}
	\end{align}
where $(\lambda_1,\dotsc,\lambda_d)$ are the eigenvalues of the Hessian matrix $\nabla\nabla\vare$
and $(q_1,\dotsc,q_d)$ are coordinates along orthogonal eigenvectors of the Hessian.  Suppose there are $m$ positive and $n$ negative eigenvalues.  The numbers $m$ and $n$ determine the shape of the constant-energy surfaces of the dispersion relation; for example, for $d=3$, these surfaces may be ellipsoids, one-sheeted hyperboloids, or two-sheeted hyperboloids.  Rescaling coordinates to 
$(p_1, \dotsc, p_n, P_1, \dotsc, P_m)$ and $\xi=\omega-\vare_0$, and using $\abs{\lambda_1\dots \lambda_d} = \abs{\det \nabla\nabla\vare(\kkk_0) } $, one obtains
	\begin{align}
	g(\omega)
	&=\frac{2^{-d/2} \pi^{-d}}{\abs{\det \nabla\nabla\vare}^{1/2} }   I_{mn}
			,\\
	I_{mn}
	&=	
		\int_{-L}^L            \!\!\!dp_1 \dots dp_n
		\int_{-\infty}^\infty  \!\!\!dP_1 \dots dP_m  ~
		\delta  (\xi + \abs{\ppp}^2 - \abs{\PPP}^2)
		.
	\end{align}
To obtain meaningful results as $\xi\rightarrow 0$, the integratiosn over $P_1,\dotsc,P_n$ should be extended to infinity, but the integrals over $p_1,\dotsc,p_m$ must be cut off at a wavenumber $L$ on the scale of the Brillouin zone.

For $\xi>0$ the leading singular terms in $I_{mn}$, excluding constant backgrounds, are
	\begin{align*}
		I_{00} &= 0
	&	I_{01} &= 0
	&	I_{02} &= 0
	&	I_{03} &= 0
	\\
		I_{10} &= \tfrac{1}{2\sqrt{\xi}} 
	& I_{11} &\approx \ln \tfrac{4L^2}{\xi}
	& I_{12} &\approx - 2\pi\sqrt{\xi}
	\\
		I_{20} &= \pi
	& I_{21} &= 0
	\\
		I_{30} &= 2\pi\sqrt{\xi}
		.
	\end{align*}
For $\xi<0$ we may use $I_{mn} (\xi) = I_{nm} (-\xi)$.

Equation~\eqref{DispersionHessianExpansion} assumed that the Hessian, $\nabla\nabla\vare$, is finite at $\kkk_0$.  For tight-binding models that exhibit flat bands where the Hessian is zero, Dirac cones where the Hessian is infinite, Weyl cones, or other unusual features in the band structure, the critical points should be treated on a case-by-case basis.

If there are two or more critical points $\kkk_1,\kkk_2,\dotsc$ contributing to a van Hove singularity at the same energy $\vare_0$, the singular forms simply combine additively:
	\begin{align}
	g^\text{sing} (\omega)
	&=\sum_{\alpha} g^\text{sing}_\alpha (\omega)
		.
	\label{CombiningSingularities}
	\end{align}

\myheading{Singularity subtraction for the DoS} 
We may exploit knowledge of the van Hove singularities as follows:

\begin{enumerate}
\item Choose an approximate DoS, $f(\omega)$, whose singularities mimic the van Hove singularities in the density of states, $g(\omega)$.
\item Calculate its Chebyshev coefficients $f_n$ analytically.
\item Subtract the approximation in the Chebyshev domain to obtain the residual Chebyshev coefficients, 
   $h_n = g_n - f_n$.
\item Construct the residual function $h(\omega)$ from the coefficients $h_n$, for certain values of $\omega$.
\item Construct $g(\omega) = f(\omega) + h(\omega)$ for these $\omega$ values.
\end{enumerate}

Since $g(\omega)$ and $f(\omega)$ have similar singularities, $g_n$ and $f_n$ will have the same tails at large $n$, and $h_n$ will decay faster.  Thus the Chebyshev series for $h(\omega)$ will converge faster.

\begin{figure*}[!bth]
		\includegraphics[width=0.99\columnwidth]{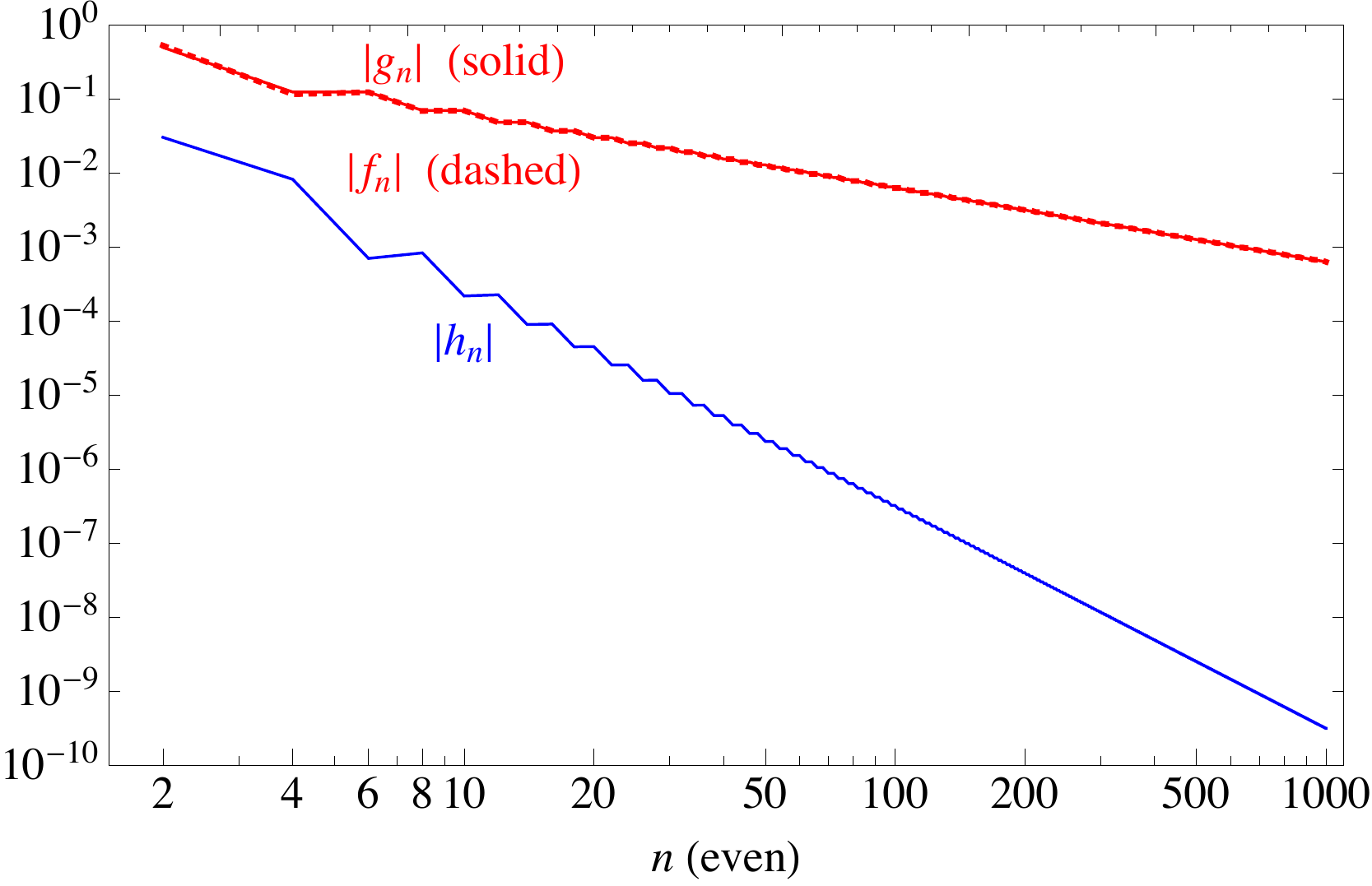}
		\includegraphics[width=0.99\columnwidth]{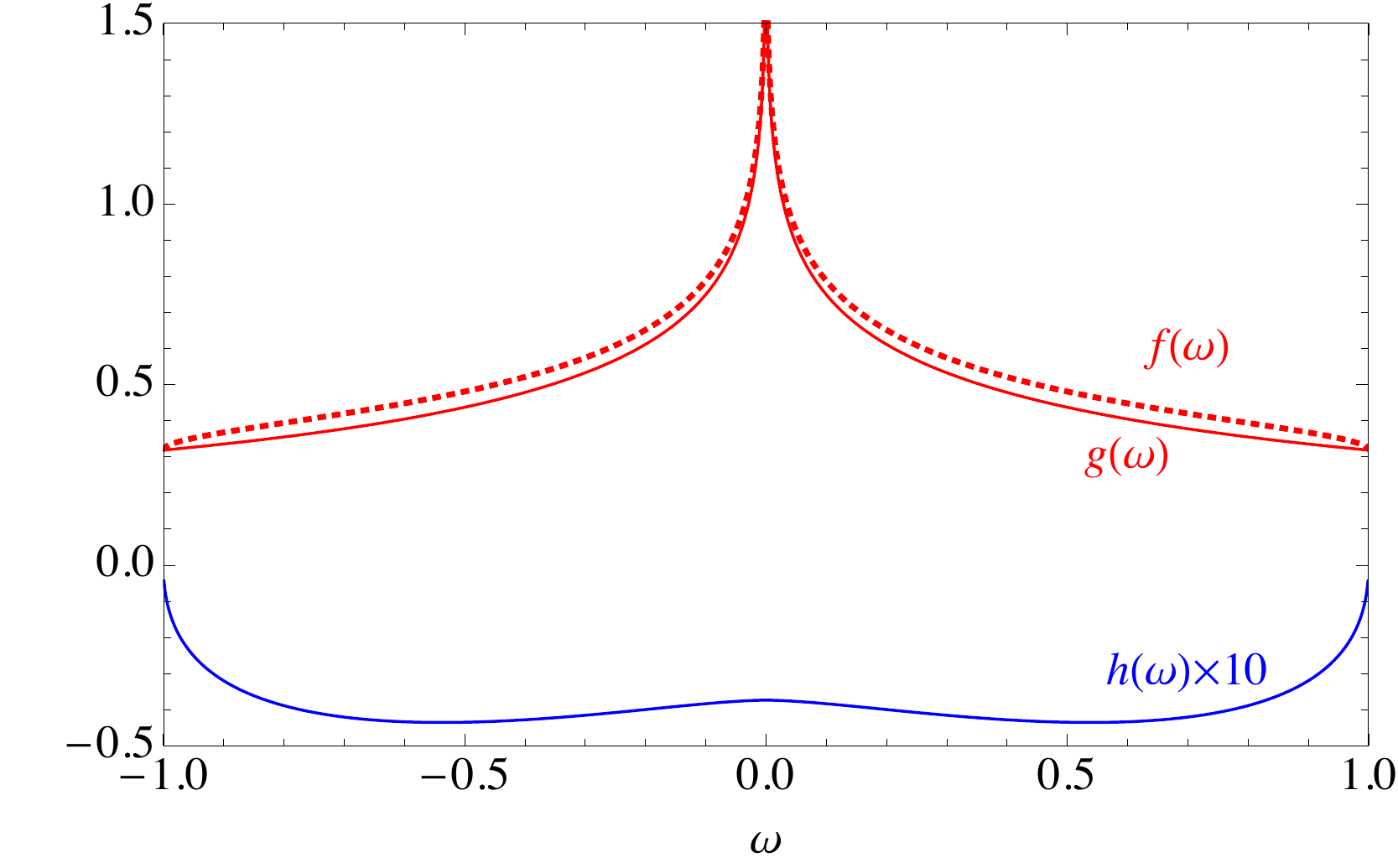}
	\caption{
	\label{SquareLattice-fgh}
	(Left)
		Chebyshev coefficients of the true square lattice density of states ($g_n$), 
		of the approximate DoS ($f_n$), 
		and of the residual ($h_n$),
		on a log-log plot.
		$g_n$ and $f_n$ both decay as $1/n$ 
		because $g(\omega)$ and $f(\omega)$ have singularities of the form $\ln\abs{\omega}$.
		In contrast, $h_n$ decays as $1/n^3$.  
		The truncation error of the series can be estimated as $\abs{h_{1000}} < 10^{-9}$.
	(Right)
		$f(\omega)$ is an approximation to the DoS that correctly captures the van Hove singularities
			at $\omega=0$ and $\omega= \pm 1$.
		$h(\omega)$ is the ``correction'' computed from the residual coefficients $h_n$;
			it has much weaker singularities.
		$g(\omega) = f(\omega) + h(\omega)$ is the corrected DoS.
	}
	\end{figure*}

\myheading{Illustration for square lattice DoS} 
The square lattice DoS has band-edge step discontinuities and a mid-band logarithmic divergence,
	\begin{align}
	g(\omega) \approx 
	\begin{cases}
		\frac{2}{\pi^2} \ln \frac{1}{\abs{\omega}} + \text{const} & \omega \approx 0	\\
		\frac{1}{\pi}	 & \omega \lesssim 1 .
	\end{cases}
	\end{align}
Construct an approximation using the functions listed in Table~\ref{ChebyshevTransforms}, and write down its Chebyshev coefficients:
	\begin{align}
	f(\omega) 
	&= 
		\frac{2}{\pi} \cdot \frac{\ln \frac{1}{\abs{\omega}}}{\pi\sqrt{1-\omega^2}}
	+	\frac{1}{\pi} \cdot 1
	,
		\\	
	f_n
	&= 
		\frac{2}{\pi} \cdot
		\left(
			\begin{cases}
			\ln 2 						& n=0	\\
			\frac{(-1)^{n/2}}{n} & n=2,4,6,\dotsc
			\end{cases}
		\right)
					\nonumber\\&~~{}
	+	\frac{1}{\pi} \cdot 
		\left(
			\begin{cases}
			\frac{2}{1-n^2}						& n=0,2,4,\dotsc	\\
			0 &
			\end{cases}
		\right)	
		.
	\end{align}
The exact Chebyshev coefficients, from Table~\ref{ChebyshevMomentFormulas}, are
	\begin{align}
	g_n
	=
			\begin{cases}
			1					& n=0	\\
	 		2^{n-1} \binom{n}{n/2} ^2
			{}_3F_2 \left(  \tfrac{n}{2}, \tfrac{n}{2}, \tfrac{n}{2}; 1-n, \half-\tfrac{n}{2}; 1 \right)
							 & n=2,4,\dotsc
			\end{cases}
		.
	\end{align}
Let $h_n=g_n-f_n$ and calculate
	\begin{align}
	h(\omega) 
	&=\frac{1}{\pi\sqrt{1-\omega^2}} \sum_{n=0}^\infty  (2-\delta_n) T_n(\omega)  h_n
	.
	\label{gSeriesImproved}
	\end{align}
The functions $f_n$, $h_n$, $g_n$, $f(\omega)$, $h(\omega)$, and $g(\omega)$ are shown in Figs.~\ref{SquareLattice-fgh}.  By dealing with the leading-order van Hove singularities analytically, we have decreased the truncation error of the 1000-term series from $10^{-3}$ to $10^{-9}$.

\myheading{Illustration for bcc lattice DoS} 
The van Hove singularities for the bcc lattice density of states are difficult to derive, because the Hessian is zero at the critical points and one has to expand the dispersion relation to third order, and because it is not clear how to cut off the logarithmic divergence correctly.  Here we ``cheat'' by using the leading terms in the series expansion of the closed form involving elliptic integrals:\cite{katsura1971,guttmann2010}
	\begin{align}
	g(\omega) \approx 
	\begin{cases}
		\frac{2}{\pi^3} \ln^2 \frac{\abs{\omega}}{8} - \frac{1}{2\pi} & \omega \approx 0	\\
		\frac{\sqrt{8}}{\pi^2} \sqrt{1-\abs{\omega}}	 & \omega \lesssim 1 .
	\end{cases}
	\end{align}
We now construct an approximation using the functions listed in Table~\ref{ChebyshevTransforms}, and write down its Chebyshev coefficients:
	\begin{align}
	f(\omega) 
	&= 
		\frac{2}{\pi^2}      \cdot \frac{\ln^2 \frac{1}{\abs{\omega}}}{\pi\sqrt{1-\omega^2}}
	+	\frac{4\ln 8}{\pi^2} \cdot \frac{\ln   \frac{1}{\abs{\omega}}}{\pi\sqrt{1-\omega^2}}
	,
		\\	
	f_n
	&= 
		\frac{2}{\pi^2} 
		\left[
			\begin{cases}
			\frac{\pi^2}{12} + \ln^2 2 						& n=0	\\
			(-1)^{n/2}
			\left(  \frac{2}{n^2} + \frac{2H_{n/2-1}+2\ln 2}{n}			\right)
				 & n=2,4,\dotsc
			\end{cases}
		\right]
					\nonumber\\&~~{}
	+	\frac{4\ln 8}{\pi^2} 
		\left[
			\begin{cases}
			\ln 2 						& n=0	\\
			\frac{(-1)^{n/2}}{n} & n=2,4,6,\dotsc
			\end{cases}
		\right]
		,
	\end{align}
where $H_n = \sum_{k=1}^n 1/k$ are the harmonic numbers.  Since $H_n \sim \ln n$ for large $n$, we see that $f_n \sim (\ln n)/n$.
The exact Chebyshev coefficients, from Table~\ref{ChebyshevMomentFormulas}, are
	\begin{align}
	g_n
	=
			\begin{cases}
			1					& n=0	\\
	 		2^{n-1} \binom{n}{n/2} ^3
		{}_4F_3 \big(  
			\tfrac{n}{2}, \tfrac{n}{2}, \tfrac{n}{2}, \tfrac{n}{2};
									\nonumber\\~~~~~~~~~~~~~~~~~~~~~
			 1-n, \tfrac{1-n}{2},  \tfrac{1-n}{2}; 1 \big)
							 & n=2,4,\dotsc .
			\end{cases}
	\end{align}
As before, we compute $h_n=g_n-f_n$ and $h(\omega)$.
The results are shown in Figs.~\ref{SquareLattice-fgh}.  Again, the singularity subtraction has decreased the truncation error of the 1000-term series from $10^{-3}$ to $10^{-9}$.

	\begin{figure*}[!bth]
		\includegraphics[width=0.99\columnwidth]{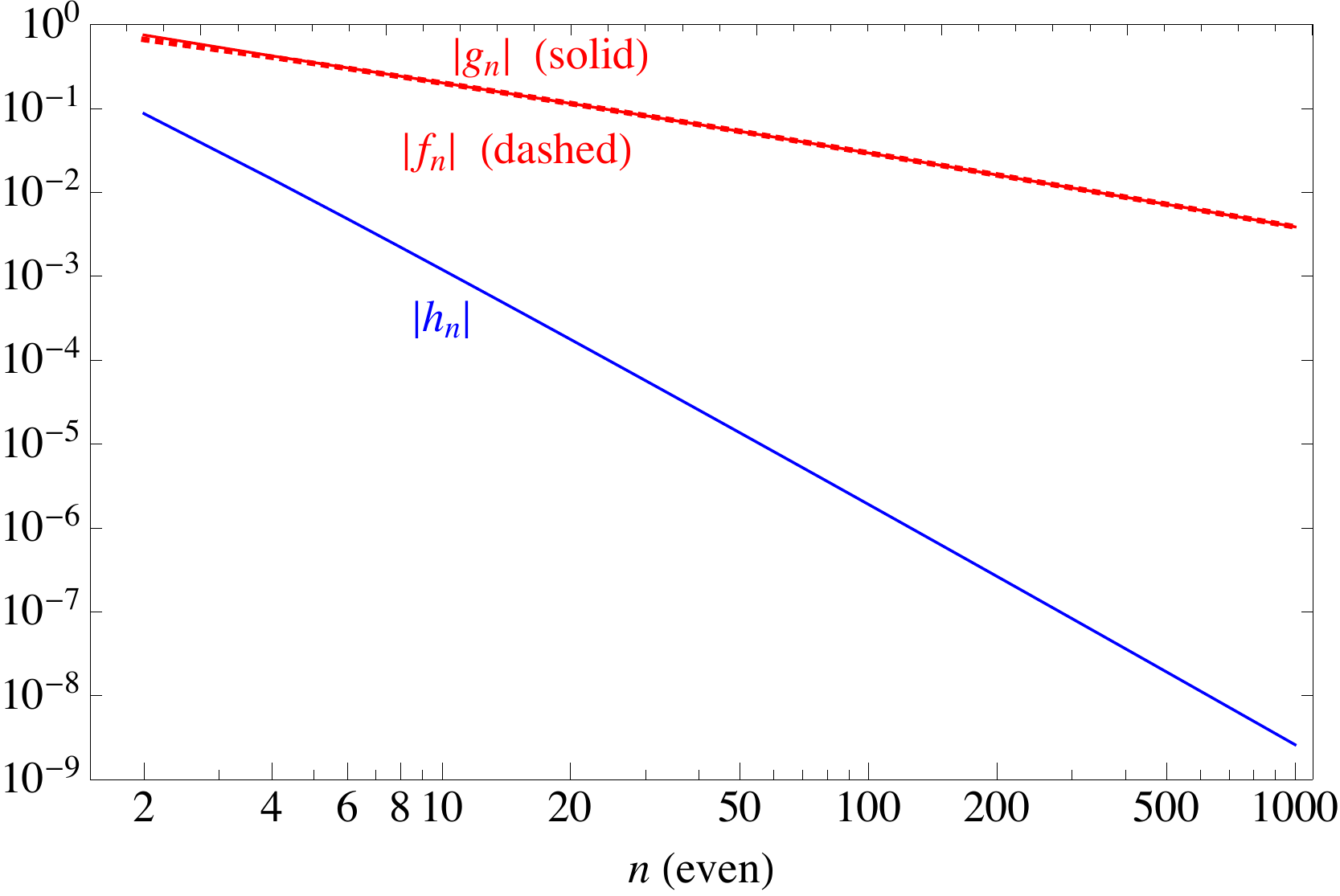}
		\includegraphics[width=0.99\columnwidth]{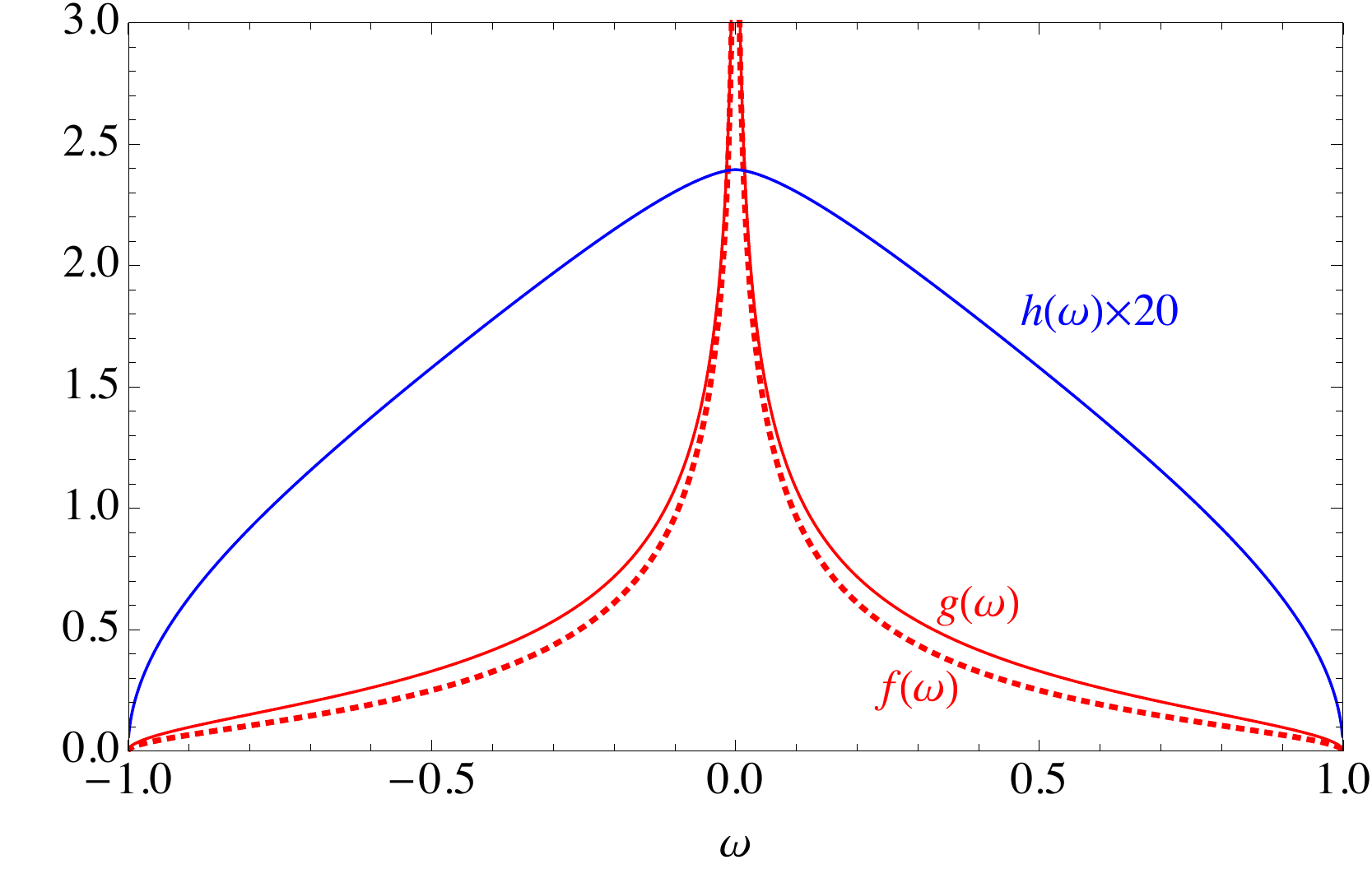}
	\caption{
	\label{BCC-fgh}
	(Left)
		Chebyshev coefficients and
	(right)
		functions 
		for the bcc lattice density of states.
		$g_n$ and $f_n$ both decay as $(\ln n)/n$ 
		because $g(\omega)$ and $f(\omega)$ have singularities of the form $\ln^2 \abs{\omega}$.
		The residual coefficients $h_n$ appear to decay as $(\ln n)/n^3$.  
	}
	\end{figure*}
	
The results agree with closed forms for $g^\text{sq} (\omega)$ and $g^\text{bcc} (\omega)$ in terms of complete elliptic integrals.\cite{guttmann2010}  

\myheading{Singularity subtraction for nonlocal spectral functions} 
We now consider the spectral function, Eq.~\eqref{spectrum}, for injecting a particle at the origin and removing it at position $\rrr$.  We have
	\begin{align}
	g_\rrr (\omega) = -\tfrac{\Im G_\rrr (\omega)}{\pi} 
	=\int_\text{BZ} \frac{d^dk~}{(2\pi)^d} ~e^{i\kkk\cdot\rrr}~ \delta \big(\omega - \vare(\kkk)\big)
	.
	\end{align}	
Van Hove singularities in $g_\rrr (\omega)$ arise from regions $\kkk~\approx~\kkk_\alpha$ where $\abs{\nabla\vare} \approx 0$.  We can generalize Eq.~\eqref{CombiningSingularities} such that the contribution from each critical point $\kkk_\alpha$ is weighted by a different phase factor:
	\begin{align}
	g^\text{sing}_\rrr (\omega)
	&=\sum_{\alpha} e^{i\kkk_\alpha \cdot \rrr} g^\text{sing}_\alpha (\omega)
		.
	\label{CombiningSingularitiesWithPhases}
	\end{align}
	
\myheading{Square lattice spectrum} 
For the square lattice, the dispersion relation $\vare(\kkk)$ has two saddle points at $(\pi,0)$ and $(0,\pi)$.  Thus the logarithmic singularity at $\omega=0$ is modified to 
	\begin{align}
	g_{xy}^\text{sing} (\omega)
	&=
		(e^{i\pi x} + e^{i\pi y}) \frac{1}{\pi^2} \ln \frac{1}{\abs{\omega}} 
		.
	\end{align}
We have verified that subtracting this singularity reduces the truncation error in $g_{xy} (\omega)$ to about $10^{-9}$, similar to the case of $g_{00} (\omega)$.

\begin{figure}[!h]
		\includegraphics[width=0.95\columnwidth]{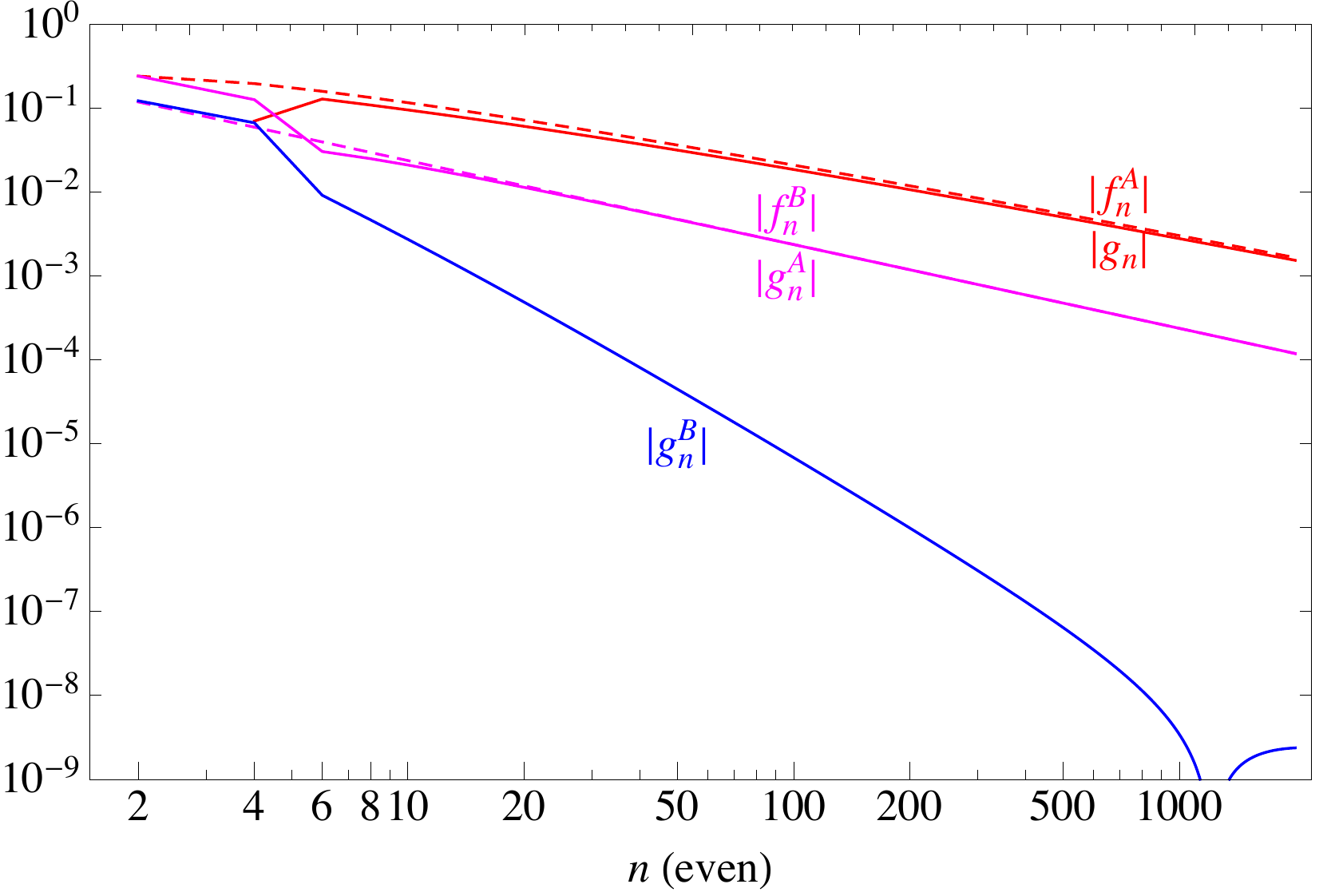}	
	\caption{
	\label{BCC-400-fghw}
		Chebyshev coefficients in the calculation of the nonlocal spectral function
		 $g^\text{bcc}_{400} (\omega)$.
		The raw Chebyshev coefficients $g_n$ and the dominant singular form $f^A_n$ 
			decay as $(\ln n)/n$.
		After subtracting the dominant singularity we are left with $g^A_n$,
			which we fit with $f^B_n$, which decays as $1/n$.
		The final residual, $g^B_n$, appears to decay as $(\ln n)/n^3$.
		Thus with 1000 terms the truncation error of the Chebyshev series is about $10^{-9}$.
	}
	\end{figure}

\myheading{BCC lattice spectrum} 
For the bcc lattice, $\vare(\kkk)$ has eight third-order saddle points at $(\pm\frac{\pi}{2},\pm\frac{\pi}{2},\pm\frac{\pi}{2})$.  The nature of the dominant $(\ln^2 \abs{\omega})$ singularities at $\omega=0$ in Fig.~\ref{BCC-Gw} can be explained in terms of Eq.~\eqref{CombiningSingularitiesWithPhases}.
Unfortunately, we are unable to predict the coefficient of the subdominant $(\ln \abs{\omega})$ singularity, because it depends delicately on the cutoff of a logarithmic integral.  
Thus we proceed as follows.  For simplicity we focus on the case where $x$, $y$, and $z$ are multiples of 4.  We know the dominant singularity to be
	\begin{align}
	f^{A} (\omega) 
	&= 
		\tfrac{2}{\pi^2}   
		\frac{\ln^2 \frac{1}{\abs{\omega}}}{\pi\sqrt{1-\omega^2}}
	,
		\\	
	f^{A}_n
	&= 
		\tfrac{2}{\pi^2} 
			\begin{cases}
			\frac{\pi^2}{12} + \ln^2 2 						& n=0	\\
			(-1)^{n/2}
			\left(  \frac{2}{n^2} + \frac{2H_{n/2-1}+2\ln 2}{n}			\right)
				 & n=2,4,\dotsc
				 .
			\end{cases}
	\end{align}
Calculate the dominant residual coefficients
	$	g^{A}_n	= g_n - f^{A}_n	$.
Assume the subdominant singularity is of the form
	\begin{align}
	f^{B} (\omega) 
	&= 
		c~
		\frac{\ln \frac{1}{\abs{\omega}}}{\pi\sqrt{1-\omega^2}}
	,
		\\	
	f^{B}_n
	&= 
		c~
			\begin{cases}
			\ln 2 						& n=0	\\
			\frac{(-1)^{n/2}}{n} & n=2,4,6,\dotsc .
			\end{cases}
	\label{FittingSubdominantSingularity}
	\end{align}
Determine the coefficient $c$ by performing a least-squares fit, over a suitable range, such that $f^{B}_n \approx g^{A}_n$.
Calculate the secondary residual coefficients
 	${	g^{B}_n	= g^A_n - f^B_n	}$.
Perform a Chebyshev transform to obtain $g^B (\omega)$.  
Finally, reconstruct the spectral function as
	\begin{align}
	g(\omega)	&\approx f^A (\omega) + f^B (\omega) + g^B (\omega).
	\end{align}
We have verified, in a few cases, that this singularity subtraction approach reduces the truncation error in $g(\omega)$ to about $10^{-9}$.  Figure~\ref{BCC-400-fghw} shows the case of $g^\text{bcc}_{400}  (\omega)$.  (According to Ref.~\onlinecite{ray2014arxiv},  $g^\text{bcc}_{xyz}  (\omega)$ is expressible in terms of hypergeometric functions.)

\myheading{Real part of Green function} 
In this section we have dealt with the DoS $g(\omega)$, which is proportional to $\Im G(\omega)$.
To obtain the complex function $G(\omega)$ we may write
$	G(\omega)	=F(\omega) + H(\omega)$
where
	\begin{align}
	H(\omega) 
	&= -\sum_{n=0}^\infty  (2-\delta_n) h_n
		\left[ U_{n-1} (\omega) + \frac{i  T_n (\omega) }{\sqrt{1-\omega^2}}   \right]
					,\\
	F(\omega) 
	&= \int_{-1}^1 d\nu~ \tfrac{1}{\omega - \nu + i 0^+} f(\nu)
		\label{F}
	.
	\end{align}
Unfortunately, for many of the singular forms for $f(\omega)$ tabulated in Table~\ref{ChebyshevTransforms}, $F(\omega)$ cannot be calculated in closed form.  
Thus Eq.~\eqref{F} may need to be evaluated numerically.

%
%

\section{Discussion} 
In this paper we calculate the Chebyshev moments combinatorially.  
Another way to obtain the moments is to calculate
$\ket{\psi_n} \equiv T_n(\hat{H}) \ket{j}$ in the site basis by repeated application of the Chebyshev recursion formula $\ket{\psi_{n+1}} = 2\hat{H} \ket{\psi_n} - \ket{\psi_{n-1}}$.  The Chebyshev moments are obtained by taking the inner product with $\bra{i}$.  This is referred to as the ``spectral method,'' ``equation-of-motion method,'' or ``kernel polynomial method'' for calculating the density of states.\cite{wangChebyshev1994,silver1994,silver1996,loh2014,karki2016}.
With this approach, computing the $L$th Chebyshev moment requires storing values of three wavefunctions on $O(L^d)$ sites, which may be memory-intensive.  The ``recursion method''\cite{berciu2010} is similar.  The continued fraction method\cite{berciu2010} may be more efficient but still requires $O(L^{d-1})$ storage.
 
In place of Chebyshev polynomials, one can use Legendre polynomials or any other family of orthogonal polynomials.  Chebyshev polynomials have the advantage that Eq.~\eqref{gSeries} can be implemented using fast Fourier transform methods. 

We have attempted to accelerate the convergence of the Chebyshev series using techniques such as Borel summation or Wynn's epsilon rule.  For a fixed value of $\omega$, we computed the partial sums of the Chebyshev series and applied convergence acceleration transformations.  We only achieved limited success.  In our opinion, it is more effective to fit the Chebyshev coefficients as in Eq.~\eqref{FittingSubdominantSingularity}, which allows us to ``accelerate'' the convergence of the series ``globally'' for all values of $\omega$ simultaneously.

We have considered other methods of analytically continuing the inverse power series into the unit disk (Fig.~\ref{Argand}).  For example, one can use the inverse power series to calculate the derivatives of the Green function, $G^{(k)} (\Omega)$  ($k=0,1,2,\dotsc,L$), at a ``pivot'' point $\Omega = \omega + i\sqrt{1-\omega^2}$ in the complex plane.  One can then calculate the Green function using the power series about the pivot point, $G(\omega) \approx \sum_{k=0}^L \frac{ G^{(k)} (\Omega) }{k!} (\omega-\Omega)^k$.
We have found that $G^{(k)} (\Omega)$ must be evaluated extremely accurately, up to very large values of $k$, in order to obtain $G(\omega)$ with modest accuracy.
The ``Chebyshev analytic continuation'' method in this paper is preferable, as it naturally lends itself to exact integer arithmetic.

In many cases $G(\omega)$ is related to Gauss, Appell, or Lauricella hypergeometric functions.\cite{guttmann2010,ray2014arxiv}  In those cases our method may be viewed as a method (albeit an indirect one) for analytic continuation of hypergeometric functions.

\section{Conclusions} 
We have developed and demonstrated a general and efficient method for calculating lattice Green functions.
The method relies on combinatorial formulas for the numbers of walks on the lattice,
which are available for bcc-like, cubic-like, and honeycomb-like lattices.
The method can be used to calculate imaginary parts (spectra) as well as real parts of the Green functions.
The basic algorithm (Sec.~\ref{secBasics}) gives Green functions to about 3 decimal places by summing Chebyshev series to 1000 terms.
Singularity subtraction (Sec.~\ref{secSingularities}) increases the accuracy to about 6--9 decimals with  little extra computational effort.
Arbitrary-precision integer arithmetic is required.
Fast cosine transforms and least-squares fitting routines may be useful 
when implementing the algorithms.

\bibliographystyle{forprl}
\bibliography{lgf,cheb}

\begin{table*}[!p]
\begin{mdframed}
	\begin{align*}
	W^\text{chain}_{2N}  &= \binom{2N}{N} \\
	W^\text{sq}_{2N} &= \binom{2N}{N}^2 \\
	W^\text{bcc}_{2N}    &=\binom{2N}{N}^3 \\
	W^\text{hon}_{2N} 
		&= \sum_{j+k+l=N} \left( \frac{N!}{j!~ k!~ l!} \right)^2
		&&= \sum_{j=0}^N \binom{N}{j}^2 \binom{2j}{j}           
		&&= {}_3F_2 \left(  \half, -N, -N; 1, 1; 4  \right)
		\\
	W^\text{diam}_{2N} 
		&= \sum_{j+k+l+m=N} \left( \frac{N!}{j!~ k!~ l!~ m!} \right)^2		
		&&= \sum_{j=0}^N \binom{N}{j}^2 \binom{2j}{j} \binom{2N-2j}{N-j}
		&&= \binom{2N}{N}  {}_4F_3 \left(  \half, -N, -N, -N; 1, 1, \half-N; 1  \right)	\\
	W^\text{cubic}_{2N} 
		&= \sum_{j+k+l=N} \frac{(2N)!}{(j!~ k!~ l!)^2}	
		&&= \binom{2N}{N} W^\text{hon}_{2N}  \\
	W^\text{hcub}_{2N} 
		&= \sum_{j+k+l+m=N} \frac{(2N)!}{(j!~ k!~ l!~ m!)^2}	
		&&= \binom{2N}{N} W^\text{diam}_{2N}  \\
	W^\text{tri}_n &= \sum_{j=0}^n \binom{n}{j} (-3)^{n-j} W^\text{hon}_{2j} \\
	W^\text{fcc}_n &= \sum_{j=0}^n \binom{n}{j} (-4)^{n-j} W^\text{diam}_{2j} 
	\end{align*}
	\begin{align*}
	W^\text{chain}_{xn}
	&=
		\binom{n}{\frac{n+x}{2}}
		&&\text{if $x+n$ even, 0 otherwise}
			\\
	W^\text{sq}_{xyzn}
	&=
		\binom{n}{\frac{n+x}{2}}
		\binom{n}{\frac{n+y}{2}}
		&&\text{if $x+y+n$ even, 0 otherwise}
			\\
	W^\text{bcc}_{xyzn}
	&=
		\binom{n}{\frac{n+x}{2}}
		\binom{n}{\frac{n+y}{2}}
		\binom{n}{\frac{n+z}{2}}
		&&\text{if $x+y+z+n$ even, 0 otherwise}
			\\
	W^\text{cubic}_{xyzn}
	&=
		\sum_{jkl}^{2(j+k+l)=n-x-y-z}
		\frac{n!} { j!~ (j+x)!~  k!~ (k+y)!~  l!~ (l+z)! }
		&&\text{if $x+y+z+n$ even, 0 otherwise}
			\\
	W^\text{hon}_{xyzn}
	&=
		\sum_{jkl}^{2(j+k+l)=\lfloor n/2 \rfloor}
		\frac{\lfloor n/2 \rfloor !} { j!~ k!~ l!}
		\frac{\lceil  n/2 \rceil  !} { (j+x)!~ (k+y)!~ (l+z)!}
		&&\text{if $x+y+z=\text{mod} (n,2)$; 0 otherwise}
			\\
	W^\text{diam}_{uvwsn}
	&=
		\sum_{jklm}^{2(j+k+l+m)=\lfloor n/2 \rfloor}
		\frac{\lfloor n/2 \rfloor !} { j!~ k!~ l!~ m!}
		\frac{\lceil  n/2 \rceil  !} { (j+u)!~ (k+v)!~ (l+w)!~ (m+s)!}
		&&\text{if $u+v+w+s=\text{mod} (n,2)$; 0 otherwise}
			\\
	W^\text{tri}_{xyzn}  &= \sum_{j=0}^n \binom{n}{j} (-3)^{n-j} W^\text{hon}_{xyz,2j} \\
	W^\text{fcc}_{uvwsn} &= \sum_{j=0}^n \binom{n}{j} (-4)^{n-j} W^\text{diam}_{uvws,2j} 		
	\end{align*}
\end{mdframed}
\caption{
	Combinatorial formulas for the numbers of walks on various lattices
	(1D chain, 2D square, 3D body-centered cubic, 2D honeycomb, 3D diamond,
	 3D cubic, 4D hypercubic, 2D triangular, 3D face-centered cubic).
	$W_n \equiv W_{000n}$ is the number of closed walks of length $n$ (starting and ending at the origin).
	$W_{xyzn}$ is the number of walks of length $n$ starting at the origin
		and ending at another point $(x,y,z)$.
	We use the convention that all lattice sites have integer coordinates.  
	The honeycomb lattice is treated as a projection of two planes of the cubic lattice
		(see Fig.~\ref{honeycomb}).
	Similarly, the diamond lattice is viewed as a projection of a subset of a hypercubic lattice.
	All lattices in the table are bipartite except for the triangular and fcc lattices.
	On bipartite lattices, all closed walks have an even number of steps.
	Many of the above formulas are from Ref.~\onlinecite{guttmann2010};
		we have arranged them to illuminate similarities and differences
		between families of lattices.
	\label{Wformulas}
}
\end{table*}

\begin{table*}[!p]
\begin{mdframed}
	\begin{align*}
	g^\text{chain}_{n}  &= \delta_n \\
	g^\text{sq}_{2N}
	  &=
	  	4^{-2N}
	  	(1+\delta_N)
			 2^{2N-1} \binom{2N}{N}^2  {}_3F_2(-N, -N, -N; 1-2N, \half-N; 1)
		\\
	g^\text{bcc}_{2N}    
		&=
	  	8^{-2N}
	  	(1+\delta_N)
			 2^{2N-1} \binom{2N}{N}^3  {}_4F_3(-N, -N, -N, -N; 1-2N, \half-N, \half-N; 1)
	\end{align*}
\end{mdframed}
	\caption{
		Formulas for Chebyshev moments $g_n=\sum_{k=0}^n a_{nk} z^{-k} W_k$ on certain lattices,
		 derived and verified with the aid of \emph{Mathematica}.
		(The above Chebyshev moments are zero for odd $n$.)
		\label{ChebyshevMomentFormulas}
	}
\end{table*}

\begin{table*}[!p]
\begin{tabular}{c|cccccccccc}
\hline\hline
$n$ 
& $W^\text{chain}_n$ 
& $W^\text{sq}_n$ 
& $W^\text{bcc}_n$ 
& $W^\text{hon}_n$ 
& $W^\text{diam}_n$  
& $W^\text{cubic}_n$ 
& $W^\text{hcub}_n$ 
& $W^\text{tri}_n$ 
& $W^\text{fcc}_n$ 
&  \\   \hline
  0 & 1 & 1 & 1 & 1 & 1 & 1 & 1 & 1 & 1 \\
 1 & 0 & 0 & 0 & 0 & 0 & 0 & 0 & 0 & 0 \\
 2 & 2 & 4 & 8 & 3 & 4 & 6 & 8 & 6 & 12 \\
 3 & 0 & 0 & 0 & 0 & 0 & 0 & 0 & 12 & 48 \\
 4 & 6 & 36 & 216 & 15 & 28 & 90 & 168 & 90 & 540 \\
 5 & 0 & 0 & 0 & 0 & 0 & 0 & 0 & 360 & 4320 \\
 6 & 20 & 400 & 8000 & 93 & 256 & 1860 & 5120 & 2040 & 42240 \\
 7 & 0 & 0 & 0 & 0 & 0 & 0 & 0 & 10080 & 403200 \\
 8 & 70 & 4900 & 343000 & 639 & 2716 & 44730 & 190120 & 54810 & 4038300 \\
 9 & 0 & 0 & 0 & 0 & 0 & 0 & 0 & 290640 & 40958400 \\
 10 & 252 & 63504 & 16003008 & 4653 & 31504 & 1172556 & 7939008 & 1588356 & 423550512 \\ 
 \hline
OEIS\# &
	A000984 & A002894 & A002897 & A002893 & A002895 & A002896 & A039699 & A002898 & A002899 \\
\hline\hline
\end{tabular}
	\caption{
		Numbers of walks of length $n$ that return to the origin on various lattices.  
		These correspond to existing entries in the Online Encyclopedia of Integer Sequences (OEIS).  
		\label{Wvalues}
	}
\end{table*}

\begin{table*}[!p]
\[
\begin{array}{c|cccccccccc}
\hline\hline
z
& 2 & 4 & 8 & 3 & 4 & 6 & 8 & 6 & 12  \\ \hline
n 
& z^n g^\text{chain}_n
& z^n g^\text{sq}_n 
& z^n g^\text{bcc}_n 
& z^n g^\text{hon}_n 
& z^n g^\text{diam}_n  
& z^n g^\text{cubic}_n
& z^n g^\text{hcub}_n
& z^n g^\text{tri}_n 
& z^n g^\text{fcc}_n
&  \\   \hline
 0 & 1 & 1 & 1 & 1 & 1 & 1 & 1 & 1 & 1 \\
 1 & 0 & 0 & 0 & 0 & 0 & 0 & 0 & 0 & 0 \\
 2 & 0 & -8 & -48 & -3 & -8 & -24 & -48 & -24 & -120 \\
 3 & 0 & 0 & 0 & 0 & 0 & 0 & 0 & 48 & 192 \\
 4 & 0 & 32 & 1728 & -15 & -32 & 288 & 1344 & 288 & 11232 \\
 5 & 0 & 0 & 0 & 0 & 0 & 0 & 0 & -2880 & -69120 \\
 6 & 0 & -512 & -79872 & 141 & 1024 & -2688 & -24576 & 3072 & -887808 \\
 7 & 0 & 0 & 0 & 0 & 0 & 0 & 0 & 64512 & 11870208 \\
 8 & 0 & 4608 & 4058112 & -1503 & -12800 & -32256 & 218112 & -400896 & 34721280 \\
 9 & 0 & 0 & 0 & 0 & 0 & 0 & 0 & -245760 & -1458585600 \\
 10 & 0 & -73728 & -216956928 & 9117 & 90112 & 2820096 & -688128 & 12496896 & 4612792320 \\
 \hline\hline
\end{array}
\]
	\caption{
		Values of local Chebyshev moments 
		$g_n=\bra{\0} T_n(\hat{H}) \ket{\0}=\sum_{k=0}^n a_{nk} z^{-k} W_k$
		 on various lattices. 
		We quote the integer-valued quantities $z^n g_n$, 
		where $z$ is the coordination number of each lattice.
		The last eight series have been submitted to the Online Encyclopedia of Integer Sequences 
		as OEIS A288454, A288455, A288456, A288457, A288458, A288459, A288460, and A288461.
		\label{ChebyshevMomentValues}
	}
\end{table*}

\begin{table*}[h]
\[
\begin{array}{cc|ccccccc}
\hline\hline
a_{nk}
  & 			&	&	 & &k  & & & \\
  &      & 0 & 1 & 2 & 3 & 4 & 5 & 6 \\   \hline
  & 0 &   1 &    &    &    &    &    &    \\
  & 1 &    & 1 &    &    &    &    &    \\
  & 2 & -1 &    & 2 &    &    &    &    \\
n & 3 &    & -3 &    & 4 &    &    &    \\
  & 4 & 1 &    & -8 &    & 8 &    &    \\
  & 5 &    & 5 &    & -20 &    & 16 &    \\
  & 6 & -1 &    & 18 &    & -48 &    & 32 \\
\hline\hline
\end{array}
\]
	\caption{
		Coefficients of powers in Chebyshev polynomials, $a_{nk}$, such that
		$T_n(x) = \sum_{k=0}^n a_{nk} x^k$.  Blank elements are zero.
		\label{ank}
	}
\end{table*}

\newcommand{\ds}{\displaystyle}

\begin{table*}[h]
\begin{alignat*}{2}
\begin{array}{c|c}
\hline\hline
                     f(\omega)    & f_n \\   \hline
                     \\
\ds
\frac{1}{\pi\sqrt{1-\omega^2}}    & \delta_n  \\ [5mm]
\ds
\frac{1}{\pi} \sqrt{1-\omega^2}   & \ds \tfrac{1}{2} \delta_n - \tfrac{1}{4} \delta_{n-2}  \\ [5mm]
1                                 & \ds \frac{2}{1-n^2}  \quad (n=0,2,4,\dotsc) \\ [5mm]
1-x^2                             & \ds \frac{12}{(1-n^2)(9-n^2)}   \quad (n=0,2,4,\dotsc) \\ [5mm]
\ds
\frac{1}{\pi\sqrt{1-\omega^2}} \ln \frac{1}{\abs{\omega}}
                                      & \begin{cases}
                                      	\ln 2  & n=0 \\
                                      	\ds \frac{(-1)^{n/2}}{n}  & n=2,4,6,\dotsc \\
                                        \end{cases} 
                                        \\  [5mm]
\ds
\frac{1}{\pi\sqrt{1-\omega^2}} \ln^2 \frac{1}{\abs{\omega}}
     & \begin{cases}
			\ds \frac{\pi^2}{12} + \ln^2 2 						& n=0	\\
			\ds (-1)^{n/2}
          \left(  \frac{2}{n^2} + \frac{2H_{n/2-1}+2\ln 2}{n}			\right)
				 & n=2,4,\dotsc
			\end{cases} 
			                                        \\  [5mm]
\\
\hline\hline
\end{array}
\end{alignat*}
  \caption{
		Chebyshev transform pairs, i.e., 
		functions $f(\omega)$ for $\omega\in [-1,1]$ and Chebyshev coefficients $f_n$ for $n=0,1,2,\dotsc$
		such that
		$f(\omega) = \frac{1}{\pi\sqrt{1-\omega^2}} \sum_{n=0}^\infty (2-\delta_n) T_n(\omega) f_n$
		and
		$f_n = \int_{-1}^1 d\omega~ T_n(\omega) f(\omega)$.
		The harmonic numbers are defined as $H_n = \sum_{k=1}^n 1/k$.
		Generally, stronger divergences in $f(\omega)$ 
			correspond to more slowly decaying tails in $f_n$.
		\label{ChebyshevTransforms}
	}
\end{table*}

\end{document}